\documentclass[11pt]{amsart}
\usepackage[a4paper, total={6in, 8in}]{geometry}
\usepackage{epsfig}
\usepackage{amsmath}
\usepackage{amssymb}
\usepackage{bbold}
\PassOptionsToPackage{hyphens}{url}\usepackage{hyperref}

\usepackage{tikz}
\usetikzlibrary{plotmarks}
\usetikzlibrary{decorations.pathreplacing}
\usepackage{pgfplots}
\pgfplotsset{width=10cm,compat=1.9}
\usepgfplotslibrary{external}
\numberwithin{equation}{section}

\allowdisplaybreaks

\usepackage[labelfont=bf]{caption}
\usepackage{setspace}
\usepackage{subcaption}
\usepackage{cleveref}
\usepackage{lscape}
\usepackage{verbatim}
\usepackage{mathrsfs} 
\usepackage{multirow}
\usepackage{bm}
\usepackage{array} 
\usepackage{color}
\usepackage{bbm}

\usepackage{graphicx}
\usepackage{wrapfig}

\newcommand{\vb}{\vspace{3.2mm}}

\usepackage[normalem]{ulem}
\allowdisplaybreaks

\begin{document}

\title{Separable mixing: the general formulation and a particular example focusing on mask efficiency}

\author[M. C. J. Bootsma, K. M. D. Chan, O. Diekmann, H. Inaba]{
  M. C. J. Bootsma$^{1,2}$,
  K. M. D. Chan$^{3,4}$,
  O. Diekmann$^{2}$ and 
  H. Inaba$^{5}$}



\begin{abstract}

The aim of this short note is twofold. We formulate the general Kermack-McKendrick epidemic model incorporating static heterogeneity and show how it simplifies to a scalar Renewal Equation (RE) when separable mixing is assumed. A key feature is that all information about the heterogeneity is encoded in one nonlinear real valued function of a real variable.
Inspired by work of R. Pastor-Satorras and C. Castellano, we next investigate mask efficiency and demonstrate that it is straightforward to rederive from the RE their main conclusion, that the best way to protect the population as a whole is to protect yourself. Thus we establish that this conclusion is robust, in the sense that it also holds outside the world of network models.

\vb


\vb

\noindent
$^{1}$Julius Centre for Health Sciences and Primary Care, University Medical Centre Utrecht, Utrecht University, Utrecht, The Netherlands
    
\noindent  
$^{2}$Department of Mathematics, Faculty of Science, Utrecht University, Utrecht, The Netherlands

\noindent  
$^{3}$Korteweg-de Vries Institute, University of Amsterdam, Amsterdam, The Netherlands

\noindent  
$^{4}$Transtrend BV, Rotterdam, The Netherlands

\noindent  
$^{5}$Faculty of Education, Tokyo Gakugei University, Koganei-shi, Tokyo, Japan
    
\noindent    \vspace{0.1cm}
    
\noindent Corresponding author: Kit Ming Danny Chan ({\tt k.m.d.chan@uva.nl}).

\end{abstract}

\maketitle

\newpage 

\section{Introduction}

 The work described below was triggered when the third author of the present paper attended the lecture of R. Pastor-Satorras during the `Workshop on Epidemic Modelling: Current Challenges' in Girona, 19-21 June 2023. This lecture reported on the models, methods and results of the paper \cite{Pastor-Satorras2022} and culminated in a powerful qualitative insight: masks that protect the wearer against infection are, also in public health perspective, more efficient than masks that, if the wearer is infectious, protect its contacts against infection. This conclusion is derived in the context of network models.
 
   Already for quite a while the present authors are working on the manuscript \cite{Bootsma2023} which aims to provide a general survey of various effects of (mainly static) heterogeneity. A natural question arose: is it possible to sustain the qualitative insight by rederiving it in the context of homogeneous mixing models? As we show below, the methodology developed in our manuscript in preparation allows us to easily provide an affirmative answer!

\section{Formulation of a comprehensive model for epidemic outbreaks in heterogeneous host populations}

   By using the word `outbreak', we imply that demographic turnover is ignored and that infection leads to permanent immunity. Host individuals are characterized by a trait $x$ taking values in a set $\Omega$. We assume that $\Omega$ is a measurable space, meaning that it comes equipped with a $\sigma$-algebra. We introduce a positive measure $\Phi$ on $\Omega$ to describe the distribution of the trait in the host population. We normalize $\Phi(\Omega)= 1$ and denote the host population size by $N$. For a concrete example see Section \ref{SectionMasks} below.

   A major restriction is that the trait of an individual does not change during the outbreak (so if the trait corresponds to age, the assumption is that the duration of the outbreak is so short, that we can ignore that individuals are becoming older while it lasts).
   Let $s(t,x)$, with $s(-\infty,x) = 1$, denote the probability that an individual with trait $x$ is susceptible at time $t$. When the NUMBER of infected individuals is small, demographic stochasticity has a large impact and cannot be ignored. Our description starts when a small FRACTION of the very large host population is infected. With an informal appeal to the Law of Large Numbers, we then also interpret $s(t,x)$ as the FRACTION of individuals with trait $x$ that is susceptible at time $t$. It follows that
\begin{equation}\label{s}
            s(t,x) = \exp \left( - \int_{-\infty}^{t}F(\tau,x)d\tau \right)
\end{equation}
with $F$ the force of infection as a function of time and trait.
   In the spirit of \cite{Kermack1927} (for a reformulation in modern language see \cite{Breda2012}) we introduce as the key modelling ingredient
   
   \begin{equation}\label{2.1}
\begin{aligned}
     A(\tau,x,\xi) =& {\rm ~ the ~expected ~contribution ~to ~the ~force ~of ~infection ~on ~an ~individual ~with ~trait ~{\it x}}\cr 
     &{\rm of ~an ~individual ~with ~trait ~\xi ~that ~became ~infected ~\tau ~units ~of ~time ~ago}
     \end{aligned}
\end{equation}

Here $A$ is a measurable non-negative function mapping $\mathbb R_+ \times \Omega \times \Omega$ into $\mathbb R_+$ and $A$ is integrable with respect to $(\tau,\xi)$ over $\mathbb R_+ \times \Omega$. 

The formula
\begin{equation}\label{2.3}
   F(t,x) =N\int_{0}^{\infty} \int_\Omega  A(\tau, x, \xi) F(t-\tau, \xi) s(t-\tau, \xi) \Phi(d\xi) d\tau,
\end{equation}
expresses the force of infection as a sum of contributions of individuals that were infected time $\tau$ ago while having trait $\xi$. By integrating \eqref{2.3} over time, interchanging the integrals, using the differentiated version of \eqref{s} to evaluate and inserting the result at the rhs of \eqref{s}, we arrive at the nonlinear abstract Renewal Equation (RE)
\begin{equation}\label{RE}
     s(t,x)=\exp\left( - N \int_{0}^{\infty} \int_\Omega  A(\tau,x,\xi) [ 1 - s(t-\tau,\xi)] \Phi(d\xi) d\tau\right)
\end{equation}

Equation \eqref{RE} provides a concise representation of a rather general class of models. For quantitative work the discrete time variant introduced in \cite{Diekmann2021} might be more suitable, especially when $\Omega$ is (or can be approximated, in some sense, by) a finite set, see \cite{Messina2022a,Messina2022b,Messina2023} for steps in this direction. 
   
An alternative way to increase the tractability is to assume separable mixing, or, in other words, to assume that $A$ is a product of a function of $x$ and a function of $(\tau,\xi)$, reflecting that the properties of the susceptible individual and the infected individual have independent influence on the likelihood of an encounter and concomitant transmission. We shall go one step further, and assume that $A$ is the product of three factors, the functions $a(x)$, $b(\tau)$ and $c(\xi)$. So also the age of infection and the trait of the infected individual are assumed to have independent influence on transmission.

\section{Separable mixing}

When
\begin{equation}\label{3.1}
   A(\tau,x,\xi)  =  a(x) b(\tau) c(\xi),
   \end{equation}
it follows straight away from \eqref{2.3} that the force of infection factorizes as a product of $a(x)$ and an unknown function of time. The same holds for the cumulative force of infection and accordingly we put
\begin{equation}\label{3.2}
  s(t,x) = e^{-a(x) w(t)},
  \end{equation}
and find that $w$ should satisfy the scalar nonlinear RE
\begin{equation}\label{3.3}
  w(t) = \int_{0}^{\infty} b(\tau) \Psi( w(t - \tau ) ) d\tau,
  \end{equation}
where $\Psi: \mathbb R \to  \mathbb R$ is defined by

\begin{equation}\label{3.4}
  \Psi(w):=  N \int_\Omega  c(\eta) ( 1 - e^{-a(\eta) w} ) \Phi(d\eta).
  \end{equation}

In the 'trivial' case that both $c$ and $a$ are identically equal to one, all individuals have identical susceptibility as well as expected infectiousness, so, after all, there is no heterogeneity. In this case 

\begin{equation}\label{3.5}
  \Psi(w) = N ( 1 - e^{-w} ),
  \end{equation}
and \eqref{3.3} is the standard Kermack-McKendrick RE as, for instance, presented in \cite{Breda2012}. So \eqref{3.3} tells us how, in the separable mixing case, the various components of heterogeneity, viz., susceptibility $a$, infectiousness $c$ and distribution $\Psi$, affect the nonlinearity in the RE. (Incidentally, in \cite{Diekmann2022}, it is shown how to efficiently derive compartmental models that incorporate heterogeneity, by choosing in \eqref{3.3} functions $b$ that are a matrix exponential sandwiched between two vectors.)

To investigate the initial phase of an outbreak, we linearize at the disease-free steady state $w=0$, which amounts to replacing $\Psi(w)$ by $\Psi’(0) w$ . Inserting the trial solution $w(t) = e^{\lambda t}$ we obtain the Euler-Lotka equation

\begin{equation}\label{3.6}
1=\Psi'(0)\int_{0}^{\infty}b(\tau)e^{-\lambda \tau}d\tau,
\end{equation}
which has a unique positive solution $\lambda = r$ whenever the Basic Reproduction Number $R_0$, given by

\begin{equation}\label{3.7}
R_0=\Psi'(0)\int_{0}^{\infty}b(\tau)d\tau,
\end{equation}
exceeds one. (The non-negativity of $b$ guarantees that in the complex plane $r$ is the right most root of \eqref{3.6}; for $R_0 < 1$ there exists a solution $r < 0$ provided the rhs of \eqref{3.6} assumes, on the real axis, values greater than one; a sufficient condition for this to happen is that $b$ has compact support). Note that 
\begin{equation}
\Psi'(0)=N\int_\Omega c(\eta)a(\eta)\Phi(d\eta).
\end{equation}

The Herd Immunity Threshold (HIT) is, by definition, reached when $w$ assumes the value $\bar{w}$ such that the reproduction number corresponding to the situation in which $\Psi’(0)$ is replaced by $\Psi’(\bar{w})$ equals one (note that after reaching the HIT there might still be a high incidence, simply because the reservoir of already infected individuals generates a considerable force of infection; but the contents of the reservoir will gradually diminish once the HIT is reached). The HIT itself is defined as $\bar{s}$, where $\bar{s}$ is the fraction of the population that is still susceptible when $w$ assumes the value $\bar{w}$. Hence

\begin{equation}\label{3.9}
\bar{s}=\int_{\Omega}e^{-a(x)\bar{w}}\Phi(dx),
\end{equation}
with $\bar{w}$ the unique (since $\Psi''(w) < 0$ ) solution of
\begin{equation}\label{3.10}
1=\Psi'(\bar{w})\int_{0}^{\infty}b(\tau)d\tau.
\end{equation}

   For $t \to \infty$,  
$w$ tends to $w(\infty)$ characterized by
\begin{equation}\label{3.11}
w(\infty)=\Psi(w(\infty))\int_{0}^{\infty}b(\tau)d\tau=\frac{\Psi(w(\infty))}{\Psi'(0)}R_0
\end{equation}
and the fraction of the population that escapes is accordingly given by
\begin{equation}\label{3.12}
s(\infty)=\int_{\Omega}e^{-a(x)w(\infty)}\Phi(dx).
\end{equation}
Note that \eqref{3.11} implies that $\Psi'(w(\infty))<\frac{\Psi'(0)}{R_0}$ (since $\Psi(y)>y\Psi'(y)$ for $y>0$) and hence that $w(\infty)>\bar{w}$.

In the next section we shall specialize the model ingredients $\Omega$, $\Phi$, $a$ and $c$ such that they reflect a situation in which a fraction $f$ of the population wears (all the time) a mask and that wearing a mask reduces, potentially, both the susceptibility and the infectiousness.

\section{Efficiency of masks}\label{SectionMasks}

Consider a population in which a fraction $f$ of the individuals wears a mask (whenever they are in a situation where they can come into contact with other individuals) while the complementary fraction $1 - f$ never wears a mask. To describe this distinction, we let $\Omega$ consist of two points, indicated by 1 and 2. We label the individuals that do not wear a mask 1 and those who do, we label 2. We specify:

\begin{align}
    \Phi(1) &= 1 - f \quad \text{and} \quad \Phi(2) = f.
\end{align}

We assume that wearing a mask is not correlated with any property that has influence on the contact process (in principle one could imagine that the contact process is assortative, in the sense that mask wearers meet disproportionately often with other mask wearers; but by this assumption we explicitly exclude such effects). Accordingly, we adopt \eqref{3.1}. Noting that this decomposition provides the freedom of incorporating multiplicative constants into the factor $b$, we normalize $a$ and $c$ by choosing:

\begin{align}
    a(1) &= 1 \quad \text{and} \quad c(1) = 1.
\end{align}

The values of $a(2)$ and $c(2)$ then describe the relative susceptibility and infectiousness of those who wear a mask. The idea that a mask offers protection is reflected in our assumption that these values lie in the interval $[0,1]$. The aim of our analysis is to investigate the influence of these values on the epidemic outbreak. Therefore we introduce parameters $\epsilon_1$ and $\epsilon_2$ and put:

\begin{align}
    a(2) &= \epsilon_1 \quad \text{and} \quad c(2) = \epsilon_2.
\end{align}

It follows that:

\begin{align}
    \Psi(w) &= N \left[(1-f)(1 - e^{-w}) + f \epsilon_2 (1 - e^{-\epsilon_1 w})\right],
\end{align}

and

\begin{align}\label{4.5}
    \Psi'(w) &= N \left[(1-f) e^{-w} + f \epsilon_1 \epsilon_2 e^{-\epsilon_1 w}\right].
\end{align}

In succession, we now consider the initial phase, the HIT and the final size, focusing on the (a)symmetry of the impact of the two parameters $\epsilon_1$ and $\epsilon_2$. As \eqref{3.6} and \eqref{3.7} show, the crucial quantities for the initial phase are $b(\tau)$ and $\Psi'(0)$. From \eqref{4.5} we deduce:

\begin{align}
    \Psi'(0) &= N \left[{ 1 - f + f \epsilon_1 \epsilon_2 }\right].
\end{align}

It follows that in the initial phase of an outbreak the two protection factors carry equal weight, in the sense that both the reproduction number $R_0$ and the Malthusian parameter $r$ depend only on their product. Motivated by this observation, we shall keep the product constant, say

\begin{align}\label{4.7}
    \epsilon_1 \epsilon_2 &= \epsilon,
\end{align}

when investigating the HIT and the final size.
\\

\textbf{Theorem 4.1:}
Assume \eqref{4.7} with $\epsilon \in (0,1)$. The HIT $\bar{s}$, defined in \eqref{3.9}, is a decreasing function of $\epsilon_1$.

\begin{proof}
Define:

\begin{align}
    G(w,\epsilon_1) &= (1-f) e^{-w} + \epsilon f e^{-\epsilon_1 w},
\end{align}

then \eqref{3.10} can be rewritten as:

\begin{align}
    G(\bar{w},\epsilon_1) &= \left(N \int_{0}^{\infty} b(\tau) d\tau \right)^{-1}.
\end{align}

Since 
\begin{align}
    \text{D}_1 G(w,\epsilon_1) & = -(1-f)e^{-w} - \epsilon_1 \epsilon f e^{-\epsilon_1 w} < 0
\end{align}
\begin{align}
    \text{D}_2 G(w,\epsilon_1) & = -w \epsilon f e^{-\epsilon_1 w} < 0 
\end{align}

we have 

\begin{align}
    \frac{d\bar{w}}{d\epsilon_1}(\epsilon_1) = -\left(\text{D}_1 G(\bar{w},\epsilon_1)\right)^{-1}\text{D}_2 G(\bar{w},\epsilon_1) < 0.
\end{align}

Next observe that the expressions for $\bar{s}$ and for $G(\bar{w},\epsilon_1)$ differ only by a factor $\epsilon$ in the last term. To exploit this, we rewrite $\text{D}_1 G \frac{d\bar{w}}{d\epsilon_1} + \text{D}_2 G = 0$ as 

\begin{align}
    -\epsilon_1 f e^{-\epsilon_1 \bar{w}} \frac{d\bar{w}}{d\epsilon_1}(\epsilon_1) - \bar{w}fe^{-\epsilon_1 \bar{w}} = \frac{1}{\epsilon} (1-f) e^{-\bar{w}} \frac{d\bar{w}}{d\epsilon_1}(\epsilon_1).    
\end{align}

Since 
\begin{align}
    \frac{d\bar{s}}{d\epsilon_1}(\epsilon_1) = -(1-f)e^{-\bar{w}} \frac{d\bar{w}}{d\epsilon_1}(\epsilon_1) - \epsilon_1 f e^{-\epsilon_1 \bar{w}}\frac{d\bar{w}}{d\epsilon_1}(\epsilon_1) - \bar{w} f e^{-\epsilon_1 \bar{w}}
\end{align}

we find

\begin{align}
    \frac{d\bar{s}}{d\epsilon_1}(\epsilon_1) = (\frac{1}{\epsilon} - 1) (1 - f) e^{-\bar{w}} \frac{d\bar{w}}{d\epsilon_1}(\epsilon_1) < 0,
\end{align}

since $0 < \epsilon < 1$.
\end{proof}

We conclude that we should minimize $\epsilon_1$ to maximize the susceptible fraction upon reaching the HIT or, in other words, we should maximize self protection.
\\

\textbf{Theorem 4.2:} Assume \eqref{4.7} with $\epsilon \in (0,1)$. The fraction $s(\infty)$ that is still susceptible after the outbreak, defined in \eqref{3.12}, is a decreasing function of $\epsilon_1$.\\

Sketch of the proof:
Define 

\begin{align}
    H(w,\epsilon_1) = (1-f) \frac{1-e^{-w}}{w} + \epsilon f \frac{1-e^{-\epsilon_1 w}}{\epsilon_1 w}
\end{align}

then \eqref{3.11} can be rewritten as the equation

\begin{align}
    H(w(\infty),\epsilon_1) = \left( N \int_0^{\infty} b(\tau) d\tau\right)^{-1}.
\end{align}

Using that $\frac{d}{dx} \frac{1-e^{-x}}{x} < 0$ for $x > 0$ one can copy the reasoning in the proof of Theorem 4.1 concerning $G$ to $H$ and derive that both $w(\infty)$ and $\bar{s}(\infty)$ are decreasing functions of $\epsilon_1$. \\

From \eqref{3.12} we have

\begin{align}
    s(\infty) = (1 - f) e^{-w(\infty)} + f e^{-\epsilon_1 w(\infty)}.
\end{align}

Since $w(\infty)$ is a decreasing function of $\epsilon_1$ the escape probability for those who do NOT wear a mask, represented by $e^{-w(\infty)}$, increases with $\epsilon_1$. From Theorem 4.2 it follows then that the escape probability of those who DO wear a mask, represented by $e^{-\epsilon_1 w(\infty)}$, decreases strongly enough to make the overall per capita escape probability $s(\infty)$ decreasing as well. 

\begin{figure}[htb]
    \centering
     \begin{subfigure}[b]{0.475\textwidth}
         \centering
         \includegraphics[height=4cm]
         {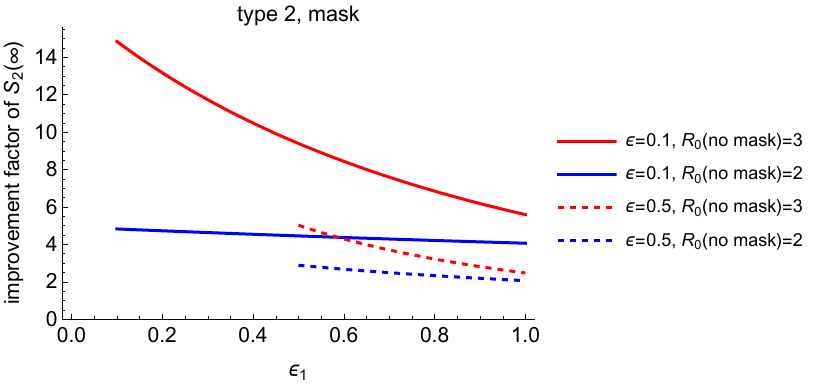}
         \caption{Individuals who always wear a mask}  
          \label{figtype2}
     \end{subfigure}
     \hfill
     \begin{subfigure}[b]{0.475\textwidth}
         \centering
         \includegraphics[height=4cm]
         {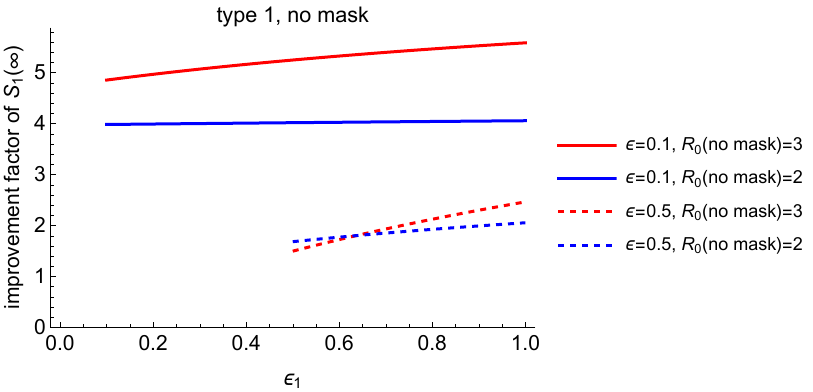}
         \caption{Individuals who never wear a mask}
         \label{figtype1}
     \end{subfigure}
  \caption{{\bf Improvement factor of escape probability for type 2 individuals who always wear a mask and for type 1 individuals who never wear a mask.}
We find the escape probabilities $s(\infty,1)$ and $s(\infty,2)$ by first numerically solving equation \eqref{3.11}. Then we compute the improvement factor of the escape probability by dividing the escape probability in a population with mask (for fraction $f$) by the escape probability in a maskless population. Curves are shown for two choices of $R_0(\text{no mask})$ and two choices of $\epsilon$, where $R_0(\text{no mask})$ is the basic reproduction number in a maskless population. Note that $\epsilon_2 = \epsilon / \epsilon_1$ as assumed in $\eqref{4.7}$.}
   \end{figure}

\begin{figure}[htb]
    \centering
     \begin{subfigure}[b]{0.475\textwidth}
         \centering
         \includegraphics[height=4cm]
         {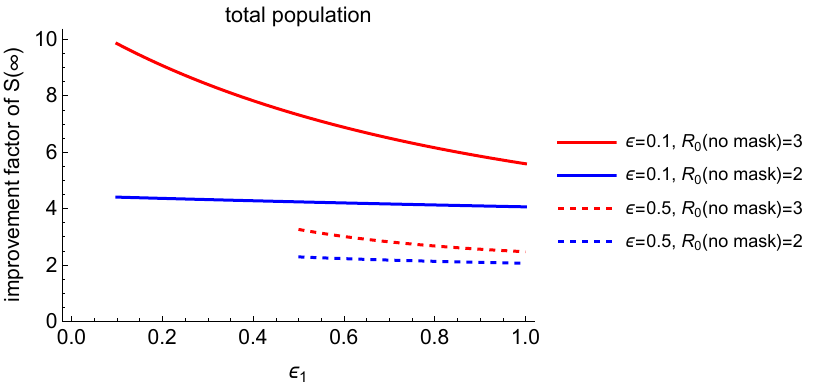}
         \caption{Escape probability for two choices of $R_0(\text{no mask})$}
         \label{figtotalr}
     \end{subfigure}
     \hfill
     \begin{subfigure}[b]{0.475\textwidth}
         \centering
         \includegraphics[height=4cm]
         {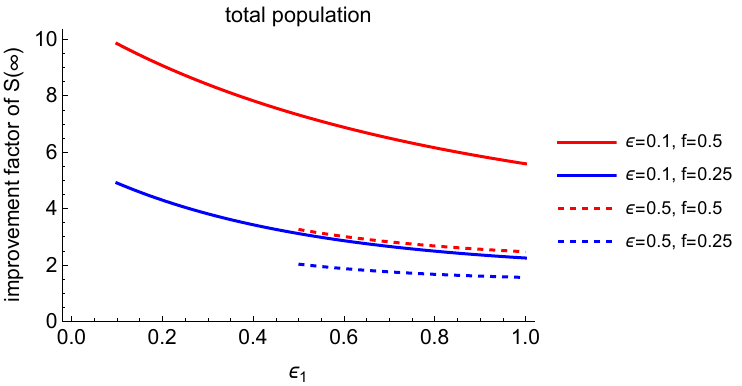}
         \caption{Escape probability for two choices of $f$}  
          \label{figtotalf}
     \end{subfigure}
   \caption{{\bf Improvement factor of escape probability for the population as a whole.}
We find the escape probability $s(\infty)$ for the population as a whole by first numerically solving equation \eqref{3.11}. The increase in escape probability is then computed by dividing the escape probability in a population with mask (for fraction $f$) by the escape probability in a maskless population. Curves are shown for two choices of $\epsilon$. In addition we show in Figure \ref{figtotalr} the impact of different choices of $R_0(\text{no mask})$: the basic reproduction number in a maskless population, while in Figure \ref{figtotalf} we show the impact of different choices of $f$. Note that $\epsilon_2 = \epsilon / \epsilon_1$ as assumed in $\eqref{4.7}$.}\label{figtotal}
   \end{figure}

Stated otherwise, maximizing self protection by those who wear a face mask improves the escape probability for themselves (Figure \ref{figtype2}) and the population as a whole (Figure \ref{figtotal}), but reduces the escape probability for those who do not wear a mask (Figure \ref{figtype1}). 

The intuitive `explanation' of the overall positive effect is that when infection of an individual is prevented, automatically the secondary infections that potentially are caused by this individual are prevented. In other words, self protection occurs one step earlier in a chain.

\section{Concluding remarks}\label{SectionRemarks}
From a strictly medical point of view, the chief aim of vaccination is to protect individuals against disease. In a public health perspective, however, one is interested in the effect of vaccination on transmission. Vaccination may reduce both the probability to get infected during an encounter with an infectious individual and the infectiousness, should infection nevertheless occur. Both reductions help to lower the force of infection and thus to diminish the size of an outbreak.

A mask is not that different from a vaccine, it too reduces both susceptibility and infectiousness. Different constructions may be more efficient in one or the other of these reductions, see \cite{Pastor-Satorras2022}. This then leads to the question of what one should strive for. In \cite{Pastor-Satorras2022} a clear conclusion is reached in the context of a SIR configuration network model (with 'random' distribution of the masks, i.e., with a form of proportionate mixing): if one keeps the product of the two reduction factors constant, one should maximize the reduction of susceptibility in order to achieve a maximal reduction of the final size.

Here we checked that the same conclusion obtains when one allows, in Kermack-McKendrick spirit, for expected infectiousness described by a general function of time elapsed since exposure and for proportionate mixing of those who do and those who do not wear a mask.

A secondary objective of the present paper is to demonstrate the effectiveness of a top down approach. Before we became aware of \cite{Pastor-Satorras2022}, we had already formulated a rather general model of an outbreak in a host population with static heterogeneity and we had studied the simplification that derives from assuming proportionate mixing. Thus the present study became, essentially, a fill in exercise.

\section*{Use of AI tools declaration}
The authors declare they have not used Artificial Intelligence (AI) tools in the creation of this article.

\section*{Acknowledgments}
It is a pleasure to thank Joan Saldaña and his team for organizing the stimulating Current Challenges Workshop on Epidemic Modelling, Girona 2023, and to thank Romualdo Pastor-Satorras for his inspiring lecture during the workshop.

\section*{Conflict of interest}

The authors declare there is no conflict of interest.

\end{document}